\documentclass[oneside,12pt,reqno]{amsart}
\usepackage{amssymb, amscd}
\usepackage[all]{xy}
\usepackage{graphics}
\usepackage{textcomp}
 \theoremstyle{definition}
 
 \theoremstyle{remark}
 
\def\noi{\noindent}

\def\la{\lambda}
\def\al{\alpha}
\def\be{\beta}
\def\ga{\gamma}

\def\BC{{\mathbb{C}}}
\def\BR{{\mathbb{R}}}
\def\BZ{{\mathbb{Z}}}
\def\BQ{{\mathbb{Q}}}

\newlength{\vscaling} \newlength{\hscaling}
\usepackage{latexsym}
\usepackage{amssymb}

\def\GG{{\mathcal G}}
\def\DD{{\mathcal D}}

\def\LL{{\mathcal L}}

\def\TT{{\mathcal T}}

\def\P2{{P^{[2]}}}
\def\vphi{{\varphi}}

\def\11{{\mbox{\boldmath $1$}}}

\def\eq{\begin{equation}}
\def\en{\end{equation}}
\def\sk{\vskip .4cm}

\def\eq#1\en{\begin{equation}#1\end{equation}}
\def\eqa#1\ena{\begin{eqnarray}#1\end{eqnarray}}

\def\hol{\mbox{hol}}
\def\#{{\sharp}}
\def\G{{\mathcal G}}
\def\Lie{\mbox{Lie}}
\def\Tr{{\mbox{Tr}}}

\newcommand{\AAA}{{\mathcal A}}

\newcommand{\HH}{\mathcal{H}}

\newcommand{\mlie}{\rm{Lie}}

\begin{document}
\begin{titlepage}
\vspace{1.cm}
\begin{flushright}
MPP-2004-89\\
LMU-TPW 04/04\\
\end{flushright}

\begin{centering}
\vspace{.43in} 
{\Large {\bf Gerbes, M5-Brane Anomalies and $\mbox{\boldmath $E_8$}$ 
Gauge Theory}\\ \vspace{1.5cm}}

\sk
{\bf Paolo Aschieri}$^{1,2,3,a}$  and ~{\bf
Branislav Jur\v co}$^{2,3,b}$\\

\vspace{.25in}$^{1}$Dipartimento di Scienze e Tecnologie Avanzate\\ 
Universit\'a del Piemonte Orientale, and INFN\\ 
Via Bellini 25/G 
15100 Alessandria, Italy\\
 \vspace{.15cm}
$^{2}$Max-Planck-Institut f\"{u}r Physik\\ F\"{o}hringer Ring 6, D-80805
M\"{u}nchen\\ \vspace{.15cm}
$^{3}$Sektion Physik, Universit\"{a}t M\"{u}nchen\\
Theresienstr. 37, D-80333 M\"{u}nchen\\ \vspace{.15cm}
\end{centering}
\vspace{.20cm}

\vspace{0.8in}

\noi {\bf Abstract}
Abelian gerbes and twisted bundles describe the topology of the NS
3-form gauge field strength $H$. We review how they  have been
usefully applied to study and resolve global anomalies in open string theory. 
Abelian 2-gerbes and twisted nonabelian gerbes describe the topology
of the 4-form field strength $G$ of M-theory. 
We show that twisted nonabelian gerbes are relevant in the study and 
resolution of global anomalies of multiple coinciding M5-branes.
Global anomalies for one M5-brane have been studied by  
Witten and by Diaconescu, Freed and Moore.
The structure and the differential geometry of twisted nonabelian 
gerbes (i.e. modules for 2-gerbes) is defined and studied. 
The nonabelian 2-form gauge potential 
living on multiple coinciding M5-branes arises as 
curving (curvature) of twisted nonabelian gerbes. The nonabelian group
is in general  $\tilde\Omega E_8$, the
central extension of the $E_8$ loop group.
The twist is in general necessary to cancel global anomalies due to the
nontriviality of the 11-dimensional 4-form  field strength $G$ and due to the
possible torsion present in the cycles the M5-branes wrap.
Our description of M5-branes global anomalies leads to the 
D4-branes one upon compactification of M-theory to Type IIA theory.  

\vspace{3.cm}

\begin{flushleft}
$^{a}$e-mail address: aschieri@theorie.physik.uni-muenchen.de\\
$^{b}$e-mail address: jurco@theorie.physik.uni-muenchen.de 
\end{flushleft}
\end{titlepage}

\setcounter{page}{1}
\section{Introduction}

The topology  of gauge theories with 2-form gauge potentials 
is a fascinating subject both from the physics and the mathematics 
perspectives. Consider for example in string theory a stack of $n$ 
coinciding D-branes, they usually form a $U(n)$ vector bundle,
however when there is a topologically nontrivial NS $B$-field
background, this is generally no more the case. Cancellation of global
anomalies requires the $U(n)$ bundle to be {\sl twisted} in order to
accommodate for the nontrivial topology of the B-field. Thus the study 
of D-brane charges in the presence of nontrivial
backgrounds leads to generalize the usual notion of fibre bundle. 
A twisted $U(n)$ bundle has transition functions $G_{ij}$ 
that satisfy the twisted cocycle relations
$G_{ij}G_{jk}G_{ki}=\la_{ijk}$, where $\la_{ijk}$ are $U(1)$ valued
functions. It
follows that $\la_{ijk}$ satisfy the cocycle relations 
$\lambda_{ijk}\lambda^{-1}_{jkl}\la_{ikl}\la_{ijl}^{-1}=1$,   
this is the characteristic property of the transition functions of 
a bundle gerbe. In short, bundle gerbes, or simply gerbes, are a higher
version of line bundles, and the gauge potential for these structures
is the 2-form $B$ in the same way as the connection 1-form $A$ is the
gauge potential associated with line bundles. As we have sketched, 
associated with a gerbe we have a twisted bundle (also
called gerbe module). 
The fact that a stack of D-branes in a nontrivial background forms 
a twisted bundle was studied in \cite{Witten:1998cd}, confirmed 
using worldsheet global anomalies in \cite{Kapustin:1999di} and 
further generalized using twisted K-theory in 
\cite{Bouwknegt:2000qt, Bouwknegt:2001vu, Carey:2002xp}.

The structure of gauge theories with 3-form gauge potentials is
similarly fascinating and rich, the corresponding geometrical structure is that of a
2-gerbe (if the 4-form field-strength is integral).
A main motivation for studying these structures is provided 
by the 3-form $C$-field of 11-dimensional supergravity. In particular it is 
interesting to study which M5-brane configurations are compatible 
with a topologically nontrivial $C$-field.
By requiring the vanishing of global anomalies, topological 
aspects of the partition function of a single M5-brane 
have been studied in \cite{Witten:1996hc, Witten:1999vg}, 
and in the presence of a nontrivial background in \cite{Witten:1999vg}
and in \cite{DFM}, \cite{Moore:2004jv}. 
We refer to \cite{Hopkins:2002rd} for the
underlying mathematical structures.

\sk

In this note we define twisted nonabelian gerbes, these are a higher
version of twisted bundles, we study their properties and show that
they are associated with abelian 2-gerbes (they are 2-gerbe modules).
Using global anomalies cancellation arguments we then see that
the geometrical structure underlying a stack of M5-branes is 
indeed that of a twisted nonabelian gerbe.
The associated 2-gerbe is constructed from the $C$-field data. The 
twist is necessary  in order to accommodate for the nontrivial topology 
of the $C$-field. A twisted nonabelian gerbe is (partly) characterized by a 
nonabelian 2-form gauge potential, in the case of a single M5-brane
this becomes the abelian chiral gauge potential of the self-dual 
3-form on the M5-brane. 
Moreover, an M5-brane becomes a D4-brane upon the appropriate 
compactification of M-theory to Type IIA string theory, and
correspondingly the 2-gerbe becomes a gerbe, and the twisted
gerbe 
becomes a twisted bundle.
A particular case is when the 2-gerbe is trivial, then the stack of
M5-branes gives a nonabelian gerbe. This corresponds in Type IIA to 
a stack of D-branes forming a bundle. The differential geometry of
nonabelian gerbes has been studied at length in \cite{Breen} 
(using algebraic geometry) and in \cite{ACJ} (using differential geometry).

It may sound strange to discuss the physical (string theory) relevance 
of twisted nonabelian gerbes before studying that of the easier case of
nonabelian gerbes.  This route is however dictated
by anomaly cancellation arguments and by the strong
analogy between M5-branes in M-theory (with open M2-branes ending on
them) and D-branes in Type IIA (with open strings ending on them). 
Indeed, as we emphasise in Section 3, the study of global open string
anomalies in the presence of a closed string NS $B$-field background
is enough to conclude that there must be a $U(1)$ gauge potential on
a $D$-brane, and that therefore a $D$-brane configuration is
associated with a line bundle. Even more, if the NS 3-form field
$H$ is torsion class (i.e. it is trivial in real De Rham cohomology but
not in integer cohomology) then we are obliged to consider coinciding
branes forming twisted $U(n)$
bundles, and this implies that $U(N)$ bundles also arise for
coinciding branes in the previous case where $B$ is torsionless.
Similarly, nontrivial backgrounds in M-theory, giving rise to
torsion classes, force us to describe the configuration of a 
stack of M5-branes via twisted nonabelian gerbes. Nonabelian gerbes are 
then recovered as a special case.

The knowledge of the topology of coinciding M5-branes is a first step
toward the formulation of the dynamics of these nonabelian gauge
fields. Indeed the full structure of a (twisted) nonabelian gerbe 
is considerably richer than just a local nonabelian 2-form gauge
potential, for example 
we also have a local 1-form gauge potential and its corresponding  
2-form field strength.
It is using all these gauge potentials and their gauge transformations
(analyzed in Section 4) that one can attack the problem of
constructing an action describing the dynamics of a stack of M5-branes.
\sk
A prominent role in nonabelian gerbes in M-theory is played by 
the $E_8$ group. Indeed, for topological considerations, the
2-form gauge potential can be always thought to be valued in
$\Omega E_8$, the $E_8$ loop group, and for twisted nonabelian gerbes in 
$\tilde\Omega E_8$, the central extension of the $E_8$ loop group.
This is so because of the simple homotopy structure of $E_8$. 
This corresponds to the fact, exploited in \cite{Evslin},
and recalled at the end of this paper, that in Type IIA theory 
a stack of D-branes gives in general a 
twisted $\tilde\Omega E_8$ bundle, so that at least for topological
considerations we can consider the gauge potential to be $\tilde\Omega E_8$
valued. 
This adds to the growing evidence that $E_8$ plays a main role in
M-theory. For example the subtle topology of the 3-form $C$-field 
is conveniently described considering it as a composite field, obtained via 
$E_8$ valued 1-form gauge potentials, roughly we have 
$C\sim CS(A_i)=\mbox{Tr}(A_i dA_i) + \frac{2}{3}\mbox{Tr}(A_i^3)$. 
Gauge theory with $E_8$ gauge group has been used in
\cite{Witten:1996md}, and then, for manifolds with boundary, in \cite{DFM} 
in order to globally define the Chern-Simons topological term 
$\Phi(C)\sim\int {1\over 6} C\wedge G\wedge G$. It has
been shown in \cite{DMW}
 to nicely confirm the K-theory formalism in Type
IIA theory upon compactification of M-theory. 
For further work in this direction see for example \cite{Evslin}, 
\cite{Sati}.
Another instance where $E_8$ gauge theory appears in M-theory is in
Ho\v rava-Witten \cite{HW}. Finally it is well known that exceptional
groups duality symmetries appear
after compactification of supergravity theories, and it has been
proposed that these symmetries follow from a hidden $E_{11}$ symmetry 
of 11-dimensional supergravity \cite{West}.

It is interesting to notice that the $E_8$ formulation 
of the $C$-field is not the only  one, in particular
in \cite{D'Auria:1982nx} another formulation related to $OSp(1|32)$ 
gauge theory was studied, and is currently investigated, 
see for example \cite{Bandos}.
It might well be that a relation between these two 
different descriptions can lead to a further
understanding of the possibly dynamical role of the $E_8$ gauge theory. 
\sk

This paper is organized as follows.
Section 2 is a review of abelian gerbes.  There are many ways of introducing
these structures (see \cite{Hitchin} for a recent and nice
introduction to the subject),  we choose a minimal approach, mainly focusing on
Deligne cohomology classes \cite{Brylinski}, these are a refinement of 
integral cohomology. Gerbes are then a geometric realizations of Deligne
classes. They are equivalent to  differential characters, also called
Cheeger-Simons characters \cite{Cheeger}, 
in this case it is the holonomy
of these higher order bundles that is emphasized.

Global worldsheet
anomalies of open strings ending on D-branes where studied in 
\cite{Freed:1999vc};
in Section 3 we use gerbes in order to construct anomaly free
worldsheet actions of strings ending on multiple coinciding D-branes.  
We mainly follow
\cite{Kapustin:1999di} and \cite{Carey:2002xp}, but also uncover some
details (especially about gauge transformations),
simplify the presentation when 
torsion is present, and emphasize that the gauge fields on the branes 
can be inferred just from the NS $B$-field in the bulk.     

Section 4 defines and studies twisted nonabelian gerbes. We 
then give an explicit construction using the loop group of $E_8$;
we also see that any twisted nonabelian gerbe can be realized by 
lifting an $E_8$ bundle.

Section 5 uses twisted nonabelian gerbes in order to describe a stack
of M5-branes.
\sk
\sk
\sk
\section{Gerbes} 

\subsection{Abelian 1-Gerbes}
Line bundles can be described using transition functions. 
Consider a cover $\{O_i\}$ of the base space $M$, then a 
line bundle is given by a set of $U(1)$ valued smooth transition 
functions $\{\la_{ij}\}$ that  satisfy $\la_{ij}=\la_{ji}^{-1}$ and
that on triple overlaps $O_{ijk}
=O_i\cap O_j\cap O_k$ satisfy the cocycle condition
\eq
\la_{ij}\la_{jk}=\la_{ik}~.
\en 
In the same spirit, a connection
on a line bundle is a set of one-forms $\{\al_i\}$ on $O_i$
such that on double overlaps $O_{ij}=O_i\cap O_j$, 
\eq\label{d1}
\al_i=\al_j+\la_{ij}d\la_{ij}^{-1}~.
\en 
Actually we are interested only in isomorphic
classes of line bundles with connection, 
indeed all physical observables are obtained
from Wilson loops, and these 
cannot distinguish between a bundle with connection 
$(\la_{ij}, \al_i)$ and an equivalent one 
$(\la'_{ij}, \al'_i)$, that by definition satisfies 
\eq\label{d2}
\la'_{ij}=\tilde\la_i \la_{ij}\tilde\la^{-1}_j~~~~~,~~~~~~
\al'_i=\al_i +\tilde\la_id\tilde\la_i^{-1}~,
\en
where $\tilde\la_i$ are $U(1)$ valued smooth functions on $O_i$.  
We are thus led to consider the class $[\la_{ij}, \al_i]$,
of all couples $(\la_{ij},\al_i)$ that satisfy (\ref{d1}), 
and where 
$(\la_{ij},\al_i)\sim(\la'_{ij},\al'_i)$ iff (\ref{d2}) holds.
The space of all these classes (called Deligne classes) is the 
Deligne cohomology group $H^1(M,\DD^1)$.
Wilson loops for the Deligne class $[\la_{ij}, \al_i]$ are given 
in Subsection 2.4.
\sk
Similarly we can consider the Deligne class 
$[\lambda_{ijk}, \al_{ij},\be_i]\in H^2(M,\DD^2)$ where now
$\lambda_{ijk}:O_{ijk}\rightarrow U(1)$ 
is totally antisymmetric in its indices, 
$\la_{ijk}=\la^{-1}_{jik}=
\la_{kij}$ etc., and it
satisfies the cocycle condition on triple overlaps
\eq\label{d31}
\lambda_{ijk}\lambda^{-1}_{jkl}\la_{ikl}\la_{ijl}^{-1}=1~,
\en
while the connection one-form $\{\al_{ij}\}$ satisfies on $O_{ijk}$
\eq\label{d32}
\al_{ij}+\al_{jk}+\al_{ki}+\la_{ijk}d\la_{ijk}^{-1}=0
\en 
and the curving two-form $\{\be_i\}$ satisfies on $O_{ij}$
\eq\label{d33}
\be_i-\be_j+d\al_{ij}=0~.
\en
The triple $(\lambda_{ijk}, \al_{ij},\be_i)$ gives
the zero Deligne class if
\eq\label{defD}
(\lambda_{ijk}, \al_{ij},\be_i)=D(\tilde\la_{ij}, \tilde\al_i)
\en
where $D$ is the Deligne coboundary operator, and $\tilde \la_{ij} :
O_{ij}\rightarrow U(1)$ are smooth functions and $\tilde\al_i$ 
are smooth one-forms on $O_i$. 
Explicitly (\ref{defD}) reads\footnote{The Deligne coboundary 
operator is $D=\pm\delta +d$, the sign factor in front of the \v Cech coboundary operator depends on the degree of the form $D$ acts on; it insures $D^2=0$.}
\eqa
\lambda_{ijk}&=&\tilde\la_{ik}\tilde\la^{-1}_{jk}\tilde\la^{-1}_{ij}\,\\
\al_{ij}&=&-\tilde\al_i+\tilde\al_j+\tilde\la_{ij}d\tilde\la^{-1}_{ij}\,,\\
\be_i&=&d\tilde\al_i\,.
\ena
There is also a geometric structure associated with the triple
$(\lambda_{ijk}, \al_{ij},\be_i)$, it is that of (abelian) gerbe 
\cite{Brylinski} or bundle gerbe \cite{Murray}. 
Equivalence classes of gerbes with connection and
curving are in 1-1 correspondence with Deligne classes, and with abuse
of language we say that $[\G]=[\lambda_{ijk}, \al_{ij},\be_i]$ is the
equivalence class of the gerbe $\G=(\lambda_{ijk}, \al_{ij},\be_i)$.  The
holonomy of an abelian gerbe is given in Subsection 2.4.  As before, gauge
invariant (physical) quantities can be obtained from the holonomy
(Wilson surface), and this depends only on the equivalence class of
the gerbe.

Gerbes are also called 1-gerbes in order to
distinguish them from 2-gerbes.
 
\subsection{Abelian 2-Gerbes}
Following the previous section, 
for the purposes of this paper, we understand under an 
abelian 2-gerbe with curvings on $M$
a quadruple $(\lambda_{ijkl},{\alpha_{ijk}, \beta_{ij},
\gamma_i})$.
Here $\lambda_{ijkl} : O_{ijkl}\equiv O_i \cap O_j \cap O_k \cap O_l
\to U(1)$ is a 2-\v Cech cocycle
\eq
\lambda_{ijkl}\lambda_{ijlm}\lambda_{jklm} =
\lambda_{iklm}\lambda_{ijkm}~~~~\mbox{ on }~ O_{ijklm}\,,
\label{cocy0}
\en
and $\lambda_{ijkl}$ is totally antisymmetric,
$\lambda_{ijkl}=\lambda_{jikl}^{-1}$ etc..
Next $\alpha_{ijk}\in \Omega^1(O_{ijk})$, 
$\beta_{ij}\in \Omega^2(O_{ij})$ and 
$\gamma_i\in \Omega^3(O_{i})$ are a collection of local one,
two, and three-forms totally antisymmetric in their respective 
indices and subject to the following relations:
\eq
\alpha_{ijk} + \alpha_{ikl} - \alpha_{ijl} - \alpha_{jkl} =
\lambda_{ijkl}d\lambda_{ijkl}^{-1}\hskip 1cm \mbox{on} \,\,\,O_{ijk}\,,\label{cocy1}
\en 
\eq
\beta_{ij}+ \beta_{jk} - \beta_{ik} = d \alpha_{ijk} \label{cocy2} 
\hskip 1cm \mbox{on} \,\,\,O_{ijk}\,,
\en
\eq
\gamma_i - \gamma_j = d \beta_{ij} \label{cocy3} \hskip 1cm \mbox{on}\,\,\, O_{ij}\,.
\en 

The equivalence class of the 2-gerbe with curvings 
$(\lambda_{ijkl},{\alpha_{ijk}, \beta_{ij},
\gamma_i})$ is given by the Deligne class
$[\lambda_{ijkl},{\alpha_{ijk}, \beta_{ij},
\gamma_i}]$, where the 
quadruple $(\lambda_{ijkl},{\alpha_{ijk}, \beta_{ij},
\gamma_i})$ represents the zero Deligne class if it is of the form
\eqa
\lambda_{ijkl} &=& \tilde\lambda^{-1}_{ijl}\tilde\lambda^{-1}_{jkl}\tilde\lambda_{ijk}\tilde\lambda_
{ikl}\,,\\
\alpha_{ijk}&=& \tilde\alpha_{ij}+\tilde\alpha_{jk}+
\tilde\alpha_{ki}+ \tilde\lambda_{ijk}d \tilde\lambda^{-1}_{ijk}\,,\\
\beta_{ij}&=&\tilde \beta_i - \tilde \beta_j + d\tilde\alpha_{ij}\,,\\
\gamma_i &=& d \tilde\beta_i\,.
\ena
The above equations are summarized in the expression
\eq
(\lambda_{ijkl},{\alpha_{ijk}, \beta_{ij},
\gamma_i})=D(\tilde\la_{ijk}, \tilde\al_{ij},\tilde\be_i)
\en
where $D$ is the Deligne coboundary operator,
$\tilde\la_{ijk}$ are $U(1)$ valued functions on $O_{ijk}$
and $\tilde\al_{ij},\tilde\be_i$ are respectively
1- and 2-forms on $O_{ij}$ and on $O_i$. 

The Deligne class $[\lambda_{ijkl},{\alpha_{ijk}, \beta_{ij},
\gamma_i}]\in H^3(M,\DD^3)$
(actually the cocycle $\{\lambda_{ijkl}\}$)
defines an integral class 
$\xi \in H^4(M,\BZ)$; 
this is the characteristic class of the 2-gerbe. 
Moreover $[\lambda_{ijkl},{\alpha_{ijk}, \beta_{ij},
\gamma_i}]$ defines the closed integral 4-form
\eq
\frac{1}{2\pi i}G = \frac{1}{2\pi i}d \gamma_i\,. \label{chern}
\en
The 4-form $G$ is a representative of
$\xi_\BR$: the image of the integral class 
$\xi$ in real de Rham cohomology.

In the same way as abelian 2-gerbes  were described above we can
define abelian $n-1$-gerbes with curvings using Deligne cohomology
classes in $H^n(M, \DD^n)$. Correspondingly we have characteristic 
classes in $H^{n+1}(M,\BZ)$.
The case $n=1$ gives equivalence classes of line bundles with
connections, and in this case the characteristic class 
is the Chern class of the line bundle. 

The relation between 
a Deligne class and its characteristic class leads to the 
following exact sequence (\cite{Brylinski}, 
see \cite{Stora} for an elementary proof)
\eq\label{sequence1}
0\rightarrow\Omega^n_{_{\BZ}}(M)\rightarrow\Omega^n(M)\rightarrow
H^n(M,\DD^n)\rightarrow H^{n+1}(M,\BZ)\rightarrow 0~
\en
where $\Omega^n_{_{\BZ}}(M)$ is the space of closed 
integral (i.e. whose integration on $n$-cycles is an integer) 
$n$-forms on $M$. 
We also have the exact sequence (see for example \cite{Gajer})
\eq\label{sequence2}
0\rightarrow H^n(M, U(1))\rightarrow
H^n(M,\DD^n)\rightarrow\Omega^{n+1}_{_{\BZ}}(M)\rightarrow 0
\en
where, as in (\ref{chern}),
$G\in \Omega^{n+1}_{_{\BZ}}(M)$ is the curvature of the
$n-1$-gerbe $(\la_{i_1,\ldots i_{n+1}}, \al_{i_1\ldots i_n},\ldots, \ga_i)$.
\sk
It is a result of \cite{Cheeger}, 
that $H^n(M, {\DD}^n)$ is isomorphic to the space of
differential characters $\check H^{n+1}(M)$ 
(Cheeger-Simons characters).  An element of 
$\check H^{n+1}(M)$ is a  pair $(h, F)$ where $h$ is a
homomorphism from the group of $n$-cycles $\BZ_{n}(M)$ to $U(1)$ 
and $F$ is an $(n+1)$-form. The pair $(h, F)$ is such that for 
any $(n+1)$-chain $\mu \in C_{n+1}(M)$ with boundary $\partial
\mu$ the following relation holds
\eq
h(\partial \mu) = \mbox{exp}(\int_\mu F)\,. \label{diffchar}
\en
The isomorphism with Deligne cohomology groups is given 
essentially via the holonomy of an $n-1$-gerbe, and $F=G$.
\sk

\subsection{Special Cases}
An important example of a 2-gerbe is obtained from an element
$\theta$ belonging to the torsion subgroup  $H^4_{tors}(M, \BZ)$
of $H^4(M, \BZ)$.
Every torsion element $\theta$ is the image of an element 
$\vartheta\in  H^3(M, \BQ/\BZ)$ via the Bockstein homomorphism 
$\be: H^3(M, \BQ/\BZ)\rightarrow H^4(M, \BZ)$ 
associated with the exact sequence $\BZ \rightarrow \BQ 
\rightarrow \BQ/\BZ$. As a \v Cech cocycle $\vartheta$ can be
represented as a $\BQ/\BZ$ valued cocycle 
$\{\vartheta_{ijkl}\}$. Now $\{\vartheta_{ijkl}\}$ can be thought of
as a \v Cech
cocycle with values in $U(1)$ valued functions on $O_{ijkl}$, 
we have of course $d\vartheta_{ijkl}=0$ and we can thus consider the 2-gerbe 
$(\vartheta_{ijkl}, 0, 0,0)\,.$
The equivalence class of this 2-gerbe is the Deligne class 
\eq\label{24}
[\vartheta_{ijkl}, 0, 0,0]\,;
\en 
it depends only on $\theta = \beta(\vartheta)$, the characteristic class 
of this Deligne class. 
\sk

Given a globally defined 3-form $C \in \Omega^3(M)$
we can construct the Deligne class 
\eq\label{trivial}
[1,0,0, C|_{O_i}]\,.
\en
Accordingly with (\ref{sequence1}) 
it has trivial characteristic class 
and it is the zero Deligne class iff 
$C \in \Omega_{_{\BZ}}^3(M)$. Indeed in this case we can write
$(1,0,0, C|_{O_i})=D(\la_{ijk},\al_{ij},\be_i)$
where $(\la_{ijk},\al_{ij},\be_i)$ is a 1-gerbe with
curvature $C$. Following \cite{Carey:2002xp},
Deligne classes like $[1,0,0, C|_{O_i}]$ will be called 
{\it trivial}. Notice that a trivial characteristic class is the same as 
a zero characteristic class while a trivial Deligne class is
usually not a zero Deligne class.
\sk
These two constructions obviously also apply to $n$-gerbes. In
particular we have the torsion 1-gerbe class 
\eq
[\vartheta_{ijk}, 0, 0]\, 
\en 
associated with the element 
$\theta\in H^3_{tors}(M,\BZ)$. 
Similarly we have the trivial 1-gerbe class
$[1,0,B|_{O_i}]$ associated with a globally defined 2-form 
$B \in \Omega^2(M)$.

\sk
Another family of 2-gerbes, the so-called Chern-Simons 2-gerbes, 
comes from a principal $G$-bundle $P_G\to M$. Its characteristic class
is the first Pontryagin class of $P_G$, $p_1 \in H^4(M, \BZ)$.
If $G$ is connected, simply
connected and simple and if $A$ is a connection on $P_G$, given locally by
a collection of $\Lie(G)$-valued one-forms $A_i$, then the image of
$p_1$
in real cohomology equals the cohomology class of $\mbox{Tr} F^2$,
and we can identify the local 
three-forms $\gamma_i$ with the Chern-Simons forms $CS(A_i)$,
$$
\gamma_i = \mbox{Tr}(A_i dA_i) + \frac{2}{3}\mbox{Tr}(A_i^3)\label{CS}\,.
$$ 
The two-forms $\beta_{ij}$ and $\alpha_{ijkl}$ and the \v Cech cocycle
$\lambda_{ijkl}$ can in principle be obtained by solving descent
equations \cite{Jouko} (see also \cite{Johnson}). 
We will denote the 2-gerbe obtained this way as $CS(p_1)$.

Notice that if
$\mbox{dim} M \leq 15$, then there is a one to one
correspondence between $H^4(M, \BZ)$ and isomorphism classes of principal $E_8$
bundles on $M$, see \cite{Witten:1985} for an elementary proof. 
This follows from the fact that the first nontrivial
homotopy group of $E_8$, except $\pi_3(E_8)=\BZ$, is
$\pi_{15}(E_8)$. We then have that up to the 14th-skeleton 
$E_8$ is homotopy equivalent to the 
Eilenberg-MacLane space $K(\BZ,3)$ (defined as the space whose only
nontrivial homotopy group is $\pi_3(K(\BZ,3))=\BZ$). Similarly
up to the 15th-skeleton 
we have $BE_8\sim K(\BZ,4)$, where $BE_8$ is the classifying space of
$E_8$ principal bundles. 
{}For the homotopy classes of
maps from $M$ to $E_8$ it then follows that
$[M,E_8]= [M, K(\BZ,3)] = H^3(M,\BZ)$  
if $\mbox{dim}M \leq 14$, and similarly $\{$Equivalence classes of $E_8$
bundles on $M \}=[M,BE_8]= [M, K(\BZ,4)]=
H^4(M, \BZ)$ if $\mbox{dim}M \leq 15$. 
Therefore, corresponding to an element $a \in H^4(M,\BZ)$ we have an $E_8$
principal bundle $P(a)\to M$ with $p_1(P(a))= a$ and picking a
connection $A$ on $P(a)$ we have a Deligne class, the Chern-Simons
2-gerbe  $CS(a)$, with $a$ being its characteristic class.

As in this paper we 
are mainly concerned with 2-gerbes associated with 5-branes embedded in
11-dimesional spacetime it is worth to recall also the homotopy 
groups of the groups $G_2$, $Spin_n$, 
$F_4$, $E_6$ and $E_7$. Except $\pi_3$ which is 
of course $\BZ$ in each case, the first nonzero ones are 
$\pi_6(G_2)$, $\pi_7(Spin_n)$ where $n\geq 7$,  $\pi_8(F_4)$, $\pi_8(E_6)$ and $\pi_{11}(E_7)$. So in the case
of a 5-brane, with 6-dimensional worldsheet $M$, we can replace $E_8$
bundles with $G_2$, $Spin_n$ where $ n\geq 7$, $F_4$, $E_6$ or $E_7$ bundles in the above discussion.

\subsection{Holonomy of Line Bundles, 1-Gerbes and  2-Gerbes}
The holonomy can be associated with any Deligne class. It gives the
corresponding differential character for cycles that arise as images of
triangulated manifolds.
Here we just collect formulas in the case of  
0-, 1- and 2-gerbes \cite{Gawedzki}, see also \cite{Carey:2002xp}.
\sk
\sk
\noi {\it Line Bundles. } 
The holonomy of $[\la_{ij}, \al_i]$ around 
a loop $\varsigma : S\rightarrow M$ can be
calculated splitting $S$ in sufficiently small arches $b$ 
and corresponding vertices
$v$, such that each $\varsigma(b)$ is completely contained in an open 
$O_i$ of the cover $\{O_i\}$ of $M$. The index $i$ depends on the arch $e$, we thus call it $\rho(e)$, and write $\varsigma(e)\subset
O_{\rho(e)}$; we also associate an index $\rho(v)$ 
with every vertex $v$ and write $\varsigma(v)\subset
O_{\rho(v)}$. We then have
\eq
\mbox{hol}(\varsigma)=\prod_{e}  \mbox{exp}{\int_e\varsigma^*\al_{\rho(e)}}
\;\prod_{v\subset e} \la^{\sigma_{e,v}}_{\rho(e)\rho(v)}(\varsigma(v))
\en
where $\sigma_{e,v}=1$ if $v$ is the final point of the oriented arch
$e$, and $-1$ if it is the initial point.
Note that the holonomy depends only on the class $[\la_{ij}, \al_i]$
and not on the representative  $(\la_{ij}, \al_i)$ or on the splitting
of $S$ or the choice
of the index map $\rho$. Of course 
if the loop is the boundary of a disk, i.e., if 
$\zeta: D\rightarrow M$ is such that $\zeta|_{\partial D}=\varsigma$,
then $\mbox{hol}(\varsigma)= e^{\int_D\zeta^*{F}}$.
\sk
\sk
\noi {\it 1-Gerbes. } 
We now consider the map $\zeta : \Sigma\rightarrow M$ where $\Sigma$ 
is a 2-cycle that we triangulate with  
faces, edges and vertices, denoted $f$, $e$ and $v$. The faces $f$
inherit the orientation of $\Sigma$, we also choose an orientation for
the edges $e$. It is always possible to choose a triangulation 
subordinate to the 
open cover $O_i$ of $M$ and define an index map $\rho$ which maps 
faces,  edges and vertices, into the index set of the covering of $M$ 
in a way that $\zeta(f) \in O_{\rho{(f)}}$, etc.. The holonomy of
the class $[\la_{ijk}, \al_{ij}, \be_i]$ is then 
\eq
\mbox{hol}(\zeta)=\prod_{f}  \mbox{exp}{\int_f\zeta^*\be_{\rho(f)}}
\,\prod_{e\subset f}  \mbox{exp}{\int_e\zeta^*\al_{\rho(f)\rho(e)}}
\,\prod_{v\subset e\subset f} 
\la_{\rho(f)\rho(e)\rho(v)}(\zeta(v))
\en
where it is understood that $\al_{\rho(f)\rho(e)}$ appears with the
opposite sign if $f$ and $e$ have incompatible orientations. Similarly
the inverse of $\la_{\rho(f)\rho(e)\rho(v)}$ appears if $f$ and $e$ have
incompatible orientations or if $v$ is not the final vertex of $e$.
As before the holonomy depends only on the  equivalence class of
the gerbe and not on the chosen representative gerbe. It is also
independent from the  choice of triangulation, 
of index map $\rho$ and of orientation of the edges. 
\sk
\sk
\noi {\it 2-Gerbes. } 
We now consider the map $\xi : \Gamma\rightarrow M$ where $\Gamma$
is a 3-cycle. We triangulate it with
tetrahedrons, faces, edges and vertices, denoted $t$, $f$, $e$ and $v$.
The triangulation is chosen to be subordinate to the
open cover $\{O_i\}$ of $M$. 
The index map $\rho$ now maps tetrahedrons, faces
etc. into the index set of the covering $\{ O_i\}$. 
The formula for the holonomy 
of the class  $[\lambda_{ijkl},{\alpha_{ijk}, \beta_{ij},
\gamma_i}]$ is 
\eqa
&&\!\!\!\!\!\!\!\!\!\!\mbox{hol}(\xi)=\\
&&\!\!\!\!\!\!\prod_t \mbox{exp}\int_t \xi^*\gamma_{\rho(t)} \prod_{f\subset t}
\mbox{exp}\int_f\xi^*\beta_{\rho(t)\rho(f)} \prod_{e\subset f\subset t}\mbox{exp}\int_e
\xi^*\alpha_{\rho(t)\rho(f)\rho(e)}\!\prod_{v \subset e\subset f\subset t}\!\lambda_{\rho(t)\rho(f)\rho(e)\rho(v)}
(\xi(v))\,.\nonumber
\ena
\sk
\sk
\sk
\section{Open strings worldsheet anomalies, 1-Gerbes and Twisted Bundles}

It is commonly said that the low energy effective action of a 
stack of $n$ branes is a $U(n)$ Yang-Mills theory. Therefore 
$n$ coinciding branes are associated with a $U(n)$ bundle. 
More in general, in the presence of a nontrivial $H$ field we 
do not have a $U(n)$ bundle, rather a twisted one, i.e. we have 
a $PU(n)$ bundle that cannot be lifted to a $U(n)$ one, 
i.e. the $PU(n)$ transition functions $g_{ij}$ cannot be
lifted to $U(n)$ transition functions $G_{ij}$ such
that under the projection $U(n)\rightarrow PU(n)$ we have 
$G_{ij}\rightarrow g_{ij}$ and such that the cocycle
condition $G_{ij}G_{jk} G_{ki}=1$ 
holds.
The twisting is necessary in order to cancel global worldsheet 
anomalies for open strings ending on D-branes.
In this section we study this mechanism.
Consider for simplicity the path integral for open bosonic 
string theory in the presence of D-branes wrapping a cycle $Q$
inside spacetime $M$ and with a given closed string background 
metric $g$ and NS three form $H$.  We have
\eq\label{openstrings}
\int\!\!\mathcal{D}\zeta~ e^{i\int_\Sigma \!\!L_{\rm{NG}}}\,e^{\int_\Sigma \zeta^* d^{-1}H\:}
\mbox{Tr$\,$hol}^{-1}_{\partial\Sigma} (\zeta^* A) 
\en
here $\zeta\,:\,\Sigma\rightarrow M$ are maps from the open 
string worldsheet $\Sigma$ to the target spacetime $M$ such that 
the image of the boundary $\partial\Sigma$ lives on $Q$, 
we denote by  $\Sigma_Q(M)$ this space,
$L_{\rm{NG}}$ is the Nambu-Goto Lagrangian, 
${\int_\Sigma \zeta^* d^{-1}H}$ is locally given by 
${\int_\Sigma \zeta^* {B}} =\int_\Sigma \varepsilon^{\al\be}
{{B}}_{MN}\partial_\al X^M\partial_\be X^N$
and is the topological coupling of the open string to the  
NS field, and Tr hol$_\gamma (\zeta^* A)$ is the trace of 
the holonomy (Wilson loop) around 
the boundary $\partial\Sigma$ of the nonabelian gauge field $A$ that lives
on the $n$ coincident D-branes wrapping $Q$. 
Now, while the exponential of the Nambu-Goto action is a well
defined function from $\Sigma_Q(M)$ to the circle $U(1)$,
the other $U(1)$ factor $e^{\int_\Sigma \zeta^*B}$ 
is more problematic because only $H=dB$ is globally defined, 
while $B=``d^{-1}H$'' is defined only locally. In order to define 
this term we need
to know not only the integral cohomology of $H$ but the full Deligne class 
$[{\mathcal G}]=[\lambda_{ijk}, \al_{ij},\be_i]$  whose curvature is $H$. 
We call the gerbe ${\mathcal G}|_Q$ trivial if its class $[{\mathcal G}|_Q]$ is trivial i.e. 
if [cf. (\ref{trivial})]:
{\it 1)} $H$ restricted to 
$Q$ is cohomologically trivial, that is it exists a $B_Q$ globally 
defined on $Q$ such that 
\eq
H|_Q=dB_Q~,
\en
and 
{\it 2)} the characteristic class $\xi$ of the gerbe is trivial  ($H|_Q$ is trivial also in integer
cohomology). It turns out that if  ${\mathcal G}|_Q$  is trivial, 
then defining
\eq\label{prod}
e^{\int_\Sigma 
\zeta^*d^{-1}H}\equiv  \hol( \Sigma\#D) 
e^{\int_{D}\tilde\zeta^* B_Q}
\en
we have a well defined function on $\Sigma_Q(M)$.
Here $D$ is the disk and $\tilde\zeta :D\rightarrow Q$ is such 
that the boundary of $\tilde\zeta(D)$ coincides with the boundary of
$\zeta(\Sigma)$ (we have assumed $Q$ simply connected and
$\partial\Sigma$ a single loop). 
Moreover $\hol( \Sigma\#D)\equiv\hol( \zeta\#\tilde\zeta)$ 
is the holonomy of the closed surface $\zeta(\Sigma)\#\tilde\zeta(D)$
obtained by gluing together $\zeta(\Sigma)$ and 
$\tilde\zeta(D)$ (and thus in particular it is obtained by 
changing the orientation of $D$).

The two terms $\hol(\Sigma\#D)$  and  $e^{\int_D \tilde\zeta^* B}$ 
depend on $\tilde\zeta : D\rightarrow Q$ and are not functions on  
$\Sigma_Q(M)$ but respectively sections of a $U(1)$ (or line) bundle 
$\partial^{-1}\mathcal{L_{[{\G}|_Q]}}^{\!\!\!\!\!\!-1}$ on $\Sigma_Q(M)$ 
and of the opposite bundle 
$\partial^{-1}\mathcal{L_{[{\G}|_Q]}}$ on $\Sigma_Q(M)$ 
so that indeed their product is 
a well defined function on $\Sigma_Q(M)$\footnote{A section of a
canonically trivial bundle such as
$\mathcal{L}^{-1}\mathcal{L}\rightarrow \Sigma_Q(M)$ 
is automatically a global function on $\Sigma_Q(M)$ because 
$\mathcal{L}^{-1}\mathcal{L}\rightarrow \Sigma_Q(M)$ 
has the canonical section $1$ (locally $1$ is the product of an 
arbitrary section $s^{-1}$ of  $\mathcal{L}^{-1}$ and of the 
corresponding section $s$ of $\mathcal{L}$) and two global sections 
define a $U(1)$ function on the base space}. 
The bundle $\partial^{-1}\mathcal{L_{[{\G}|_Q]}}\rightarrow \Sigma_Q(M)$ is
constructed from the 1-gerbe class.
Without entering this construction [described after eq.
(\ref{torsionclass})] we can 
directly see that expression (\ref{prod}) is a well defined 
function on $\Sigma_Q(M)$ by showing its independence from the 
choice of the map $\tilde\zeta$.  
Given another map $\tilde\zeta'$ we have
\eq
\hol(\zeta\#\tilde\zeta)/ \hol(\zeta\#\tilde\zeta')=
\hol(\tilde\zeta'\#\tilde\zeta)=
e^{\int_D \tilde\zeta'^* B_Q -\int_D  \tilde\zeta^* B_Q}
\en
where the first equality is the holonomy gluing property and the 
last equality holds because the integral of $B_Q$ on 
$\tilde\zeta'\#\tilde\zeta$ equals the holonomy of the 
gerbe since $B_Q$ gives a gerbe $(1,0,B_Q)$ on $Q$  equivalent
to ${\mathcal G}|_Q$: $[1,0,B_Q]=[{\mathcal G}|_Q]$.

Expression (\ref{prod}) depends on the  equivalence class of
the initial gerbe ${\mathcal G}$ and also on $B_Q$, not just on 
$[1,0,B_Q]$.
Had we chosen a different $2$-form $B'_Q$ such that 
$ [1,0,B'_Q]=[{\mathcal G}|_Q]=[1,0,B_Q]$,
then the result would have differed by the phase 
\eq\label{phase}
e^{\int_{D}\tilde\zeta^* (B'_Q-B_Q)}~,
\en
where ${1\over {2\pi i}}\omega\equiv {1\over {2\pi i}}(B'_Q-B_Q)$ 
is a closed integral $2$-form, recall (\ref{sequence2}).
In order to absorb this extra phase (this gauge transformation)
we have to consider the last term in  (\ref{openstrings}):
$\mbox{Tr hol}_{\partial\Sigma} (\zeta^* A)$. This is a well defined 
$U(1)$-valued function on $\Sigma_Q(M)$ and $A$ is a true $U(n)$ 
connection on a nonabelian bundle on $Q$, with Tr the trace in the
fundamental of $U(n)$. Under the
gauge transformation $B_Q\rightarrow B'_Q=B_Q+\omega$ 
we have to transform accordingly the $U(n)$ bundle in order to 
compensate for the phase
factor (\ref{phase}). This is obtained considering the new $U(n)$
bundle with curvature $F'=F+\omega$ obtained by tensoring the initial
$U(N)$ bundle on $Q$ with the $U(1)$ bundle on $Q$ 
defined by the closed $2$-form $\omega$
(the definition of this $U(1)$ bundle is unique since we have 
considered $Q$ simply connected). If we consider just one D-brane
we recover the gauge invariance of the total $U(1)$
field $B_Q-F$; the gauge transformations locally read 
$B_Q\rightarrow B_Q+d\Lambda$ and $A\rightarrow A+\Lambda$. 

In conclusion using anomaly cancellation we have seen that 
if the open strings couple to the $B$ field, then their ends
must couple to a $U(1)$ gauge field $A$. So far there is no
requirement for nonabelian gauge fields.
\sk

The situation is  more involved if $\G|_Q$ has torsion,
i.e. if the three form $H$ restricted to $Q$ is cohomologically trivial, but 
the characteristic class of the gerbe is nontrivial.
In this case
(\ref{prod}) is not well defined because $[1,0,B_Q]\not= [\G|_Q]$.
However any torsion gerbe can be obtained form a 
lifting gerbe, i.e. from a gerbe that describes the 
obstruction of lifting a $PU(n)$ bundle to a $U(n)$ one 
(with appropriate $n$).
We now describe this lifting  gerbe and the associated twisted $U(n)$ bundle.
Let $P\rightarrow M$ be a $PU(n)$ bundle and consider the
exact sequence $U(1)\rightarrow U(n){\stackrel{\pi}{\to}} 
PU(n)$. Consider an open cover $\{ U_\al\}$ of $PU(n)$ with sections 
$s^\al:U_\al\subset PU(n)\rightarrow U(n)$. 
Consider also a good cover $\{O_i\}$ of $M$ such that
each  transition function ${g_{ij}}$ of $P\rightarrow M$ 
has image contained in a $U_i$ (this is always doable, we also
fix a map from the couples of indices  $(i,j)$ 
to the $\al$ indices).
Let $G_{ij}=s^\al(g_{ij})$, these are $U(n)$ valued functions
and satisfy:
$$
G_{ik}G^{-1}_{jk}G^{-1}_{ij}=\la_{ijk} 
$$
where $\la_{ijk}$ is $U(1)$ valued as is easily seen applying
the projection $\pi$ and using the cocycle relation for the 
$g_{ij}$ transition functions.
We say that $G_{ij}$ are the transition functions for a $U(n)$ 
twisted bundle and that the 
lifting  gerbe is defined by the twist  $\la_{ijk}$.
It is indeed easy to check that the $\la_{ijk}$ satisfy the
cocycle condition on quadruple overlaps $O_{ijkl  }$.
A connection for a twisted bundle is a set of 1-forms $A_i$
such that $\al_{ij}\equiv -A_i+G_{ij} A_j G^{-1}_{ij}+
G_{ij}dG_{ij}^{-1}$ is a connection for the corresponding
gerbe (in particular  $\pi_*A$ is a connection on the initial
$PU(n)$ bundle $P$).
We restate this construction this way: consider the couple $(G_{ij},A_i)$,
and define
\eqa
{\mathbf D} (G_{ij},A_i)&\equiv&\left( (\delta G)_{ijk},
(\delta A)_{ij}+G_{ij}dG_{ij}^{-1},  \mbox{$1\over n$}TrdA_i
\right)\nonumber\\
&=&(G_{ik}G^{-1}_{jk}G^{-1}_{ij}  
, -A_i+G_{ij} A_j G^{-1}_{ij}+
G_{ij}dG_{ij}^{-1}, 
 \mbox{$1\over n$}TrdA_i)
\label{delignefrombundle}\,.
\ena
  If this triple has abelian entries then it 
defines a gerbe,  and $(G_{ij},A_i)$ is called a twisted bundle. We
also say that the twisted bundle $(G_{ij},A_i)$ is  twisted by the gerbe 
${\mathbf D} (G_{ij},A_i)$.
Notice that the nonabelian $\mathbf D$ operation becomes the abelian Deligne coboundary operator $D$ 
if $n=1$ in $U(n)$ [cf.(\ref{defD})].

More in general,  if 
$\G|_Q =(\la_{ijk},\al_{ij},\be_i)|_Q$,
is torsion then it follows from the results in \cite{G} that one can 
always 
find a twisted  bundle  $(G_{ij},A_i)$ such that
\eq\label{torsionclass}
(\la_{ijk}, \al_{ij},\be_i)|_Q={\mathbf D} (G_{ij},A_i) 
+ (1,0,B_Q)
\en
where $B_Q$ is a globally defined abelian 2-form.
\sk

We can now correctly define the path integral (\ref{openstrings}).
We proceed is three steps. 

{\it i)$~$} Using the holonomy gluing property it is easy to see that 
hol$(\Sigma\#D)\equiv\hol(\zeta\#\tilde\zeta)$ is a section of the line bundle
$\partial^{-1} \LL_{[-\G|_Q]}\rightarrow \Sigma_Q(M)$ at the point
$\zeta\in \Sigma_Q(M)$. The line bundle $\partial^{-1}
\LL_{[-\G|_Q]}\rightarrow \Sigma_Q(M)$ is the 
pull back to  $\Sigma_Q(M)$ of the line  bundle on loop space  
$\LL_{[-\G|_Q]}\rightarrow L(Q)$. We characterize $\LL_{[-\G|_Q]}\rightarrow
L(Q)$ (here $-\G|_Q$ is a generic gerbe over $Q$) by realizing its
sections $L(Q)\rightarrow \LL_{[-\G|_Q]}$ through functions 
$s:D(Q)\rightarrow \BC$ where $D(Q)$ is the space of maps from the
disk $D$ into $Q$; the boundaries of these maps are loops in $L(Q)$.
The function $s$ is a section of $\LL_{[-\G|_Q]}\rightarrow
L(Q)$ if $s(\tilde\zeta)=\hol(\tilde\zeta\# \tilde\zeta')s(\tilde\zeta')$
for all $\tilde\zeta\,,\tilde\zeta'\in D(Q)$ that are equal on the 
boundary:  $\tilde\zeta|_{\partial D}=\tilde\zeta'|_{\partial D}$. 
Expression $\hol(\tilde\zeta\#\tilde\zeta')$ above is the holonomy of 
$[-\G|_Q]$ on the closed surface $\tilde\zeta(D)\#\tilde\zeta'(D)$
obtained by gluing together $\tilde\zeta(D)$ and $\tilde\zeta'(D)$.

{\it ii)$~$} If we define $\TT={\mathbf D} (G_{ij},A_i)$, then (\ref{torsionclass}) reads
$\GG|_Q-{\TT}= (1,0,B_Q)$ and we see that $e^{\int_D
  \tilde\zeta^*B_Q}$ is a section
of the line bundle $\LL_{{[\GG|_Q}-\TT]}\rightarrow L(Q)$. From {\it i)$~$} and {\it ii)$~$} we see that we need a section of the line bundle 
$\LL_{[\TT]}\rightarrow L(Q)$.

{\it iii)$~$}
A section
of the line bundle  $\LL_{[\TT]}\rightarrow L(Q)$ is given by the inverse of the 
trace of the holonomy of the twisted $U(n)$ bundle 
$(G_{ij},A_i)$.
The definition is as follows.
The pull back of $\TT$ on the disk $D$ via 
$\tilde\zeta:D\rightarrow Q$ is trivial since $D$ is two dimensional.
We can thus write
\eq
{\mathbf D}(\tilde\zeta^* G_{ij}, \tilde\zeta^*
A_i)=D( \tilde\lambda_{ij},\tilde \al_i)+ (1,0,b)
\en
so that  $(\tilde\zeta^* G_{ij\,}\lambda^{-1}_{ij\,},
\tilde\zeta^*A_i-\tilde\al_i)$ is a true $U(n)$ bundle.
We then define
\eq\label{twistedholonomy}
\mbox{Tr$\,$hol}_{\partial\Sigma} (\zeta^* A) \equiv
\mbox{Tr$\,$hol}_{\partial\Sigma} (\tilde\zeta^* A-\tilde\al) 
e^{-\int_D b} 
\en
where $\mbox{Tr$\,$hol}_{\partial\Sigma} (\tilde\zeta^*A- \tilde\al)$
is the trace of the holonomy of the $U(N)$ bundle
$(\tilde\zeta^* G_{ij\,}\tilde\lambda^{-1}_{ij\,},
\tilde\zeta^*A_i-\tilde\al_i)$.
Note that if we consider the couple $(G_{ij},A_i)$
in the adjoint representation, then it defines a true $SU(n)$ 
bundle.  Consistently, if in (\ref{twistedholonomy}) we consider 
the trace in the adjoint representation
instead of the trace in the fundamental, 
we then obtain the holonomy of this $SU(n)$ bundle.
It is easy to check that definition (\ref{twistedholonomy}) 
 is independent from 
the choice of the trivialization $\tilde\lambda,\tilde \al$, $ b$ and
of the map  $\tilde\zeta:D\rightarrow Q$.
\sk
We conclude that expression (\ref{openstrings}) is well defined
because we have the product of the three sections 
\eq\label{3sections}
\mbox{hol}(\Sigma\#D)\; e^{\int_D\tilde\zeta^*B_Q}\;
\mbox{Tr$\,$hol}^{-1}_{\partial\Sigma}(\tilde\zeta^*A)~,
\en
respectively sections of  the bundles $\partial^{-1}\LL_{[-\G|_Q]}$,
$\partial^{-1}\LL_{[\G|_Q-\TT]}$ and $\partial^{-1}\LL_{[\TT]}$
on the base space $\Sigma_Q(M)$ obtained by pulling back 
the corresponding bundles on the loop space $L(Q)$ via the map
 $\Sigma_Q(M)\stackrel{\partial}{\to}L(Q)$.
The product of these three bundles is canonically trivial.
Expression (\ref{3sections}) depends on the Deligne classes $[\G]$, $[\TT]$ 
and on the potential $B_Q$; it is easily seen that it does not depend on 
the choice of the map  $\tilde\zeta:D\rightarrow Q$. In order to obtain 
a gauge invariant action the gauge transformation 
$B_Q\rightarrow B'_Q=B_Q+\omega$ comes always together with the
transformation of the twisted $U(N)$ bundle $(G_{ik}, A_i)$ obtained  
 by tensoring $(G_{ik}, A_i)$ with the $U(1)$ bundle on $Q$ 
defined by the closed $2$-form $\omega$.

In conclusion using anomaly cancellation we have seen that 
if the open strings couple to the $B$ field --more precisely 
to the gerbe class $[\G]$-- then their ends must couple to a twisted 
$U(n)$ gauge field $A$ if on the boundary $\G$ is torsion.

\sk
For sake of simplicity, up to now we have omitted spinor fields.
In superstring theory, due to the determinant of the spinor fields, 
we have an extra term entering the functional integral: Pfaff. 
This is a section of the bundle 
$\partial(\LL_{[\omega_{ijk}, 0,0]})\rightarrow \Sigma_Q(M)$ 
where $[\omega_{ijk},0,0]$ is the Deligne class associated with
the second Stiefel-Whitney class $\omega_2\in H^2(Q,\BZ_2)$ of the 
normal bundle of $Q$ [or, which is the same, with its image $W_3$ in 
$H^3_{tors}(Q,\BZ)$]. In this case we consider 
a $PU(n)$ bundle 
$P\rightarrow Q$ with curvature two form such that
instead of (\ref{torsionclass}) the following equation holds,
$
(\la_{ijk},\al_{ij}, \be_i)|_Q-(\omega_{ijk},0,0)={\mathbf D} (G_{ij},A_i) 
+ (1,0,B_Q)\,.
$
Correspondingly, the product 
\eq
\mbox{Pfaff}\,\; \mbox{hol}(\Sigma\#D)\; e^{\int_D\tilde\zeta^*B_Q}\;
\mbox{Tr$\,$hol}^{-1}_{\partial\Sigma}(\tilde\zeta^*A)~,
\en
is a well defined function on $\Sigma_Q(M)$ because  
$\partial^{-1}\LL_{[\omega_{ijk},0,0]}\,\partial^{-1}\LL_{[-\G|_Q]}
\partial^{-1}\LL_{[1,0,B|_Q]}\,\partial^{-1}\LL_{[\TT]}$ is the
trivial bundle. We thus arrive at the general 
condition for a stack of D-branes to be wrapping a cycle $Q$ in $M$. It is 
\eq\label{abcde}
[\la_{ijk},\al_{ij}, \be_i]|_Q-[\omega_{ijk},0,0]=[{\mathbf D} (G_{ij},A_i)] 
+ [1,0,B_Q]\,,
\en 
i.e. the stack of D-branes must form a twisted bundle, the twist being
given by a gerbe that up to a trivial gerbe is equal to the initial gerbe
associated with the 3-form $H$ minus the gerbe obtained from the second
Stiefel-Whitney class of the normal bundle of $Q$.  
In particular, for the characteristic classes of these gerbes we have,
\eq
[H]|_Q-W_3=\xi_{[{\mathbf D} (G_{ij},A_i)]}\,,
\label{HW3}
\en
where $[H]|_Q\equiv\xi_{\G|_Q}$ is the characteristic class of the
restriction to $Q$ of the gerbe $\G=(\la_{ijk},\al_{ij}, \be_i)$
associated with the 3-form $H$, and $W_3=\be(\omega_2)$ is the obstruction for 
having Spin$^{c}$ structure on the normal bundle of $Q$, 
in fact $\be$ is the Bockstein homomorphism associated with the 
short exact sequence
$
\BZ \stackrel{\times 2}{\rightarrow} \BZ \rightarrow \BZ_2\,.
$
\sk
\sk
\sk
\section{Twisted Nonabelian Gerbes (2-gerbe Modules)}
In this section we slightly generalize 
the notion of twisted  bundle (1-gerbe module) 
and then consider the one degree higher case. 
In (\ref{delignefrombundle}) 
twisted $U(n)$ bundles where defined. More generally,
consider the central extension:
\eq\label{central}
U(1)\to G\stackrel{\pi}{\to} G/U(1)\,
\en
i.e. where $U(1)$ is mapped into the center $Z(G)$ of $G$. (In the
following we will not distinguish between $U(1)$ and its image 
ker$\pi\subset G$).
A twisted $G$ bundle with connection $A$ and curvature $F$ is a triple 
$(G_{ij}, A_i, F_i)$ such that 
\eq
F_i=G_{ij}F_jG^{-1}_{ij}~,\label{FGFG}
\en
and such that
\eqa
{\mathbf D}_F (G_{ij},A_i)&\equiv&\left( (\delta G)_{ijk},
(\delta A)_{ij}+G_{ij}dG_{ij}^{-1}, dA_i+A_i\wedge A_i -F_i
\right)\nonumber\\
&=&(G_{ik}G^{-1}_{jk}G^{-1}_{ij}  
, -A_i+G_{ij} A_j G^{-1}_{ij}+
G_{ij}dG_{ij}^{-1}, dA_i+A_i\wedge A_i -F_i)
\label{delignefrombundle2}\,
\ena
has  $U(1)$- and $\Lie(U(1))$-valued entries.

It is not difficult to check that the triple 
(\ref{delignefrombundle2}) defines a gerbe [hint:
since the group extension is central,
$d ((\delta A)_{ij}+G_{ij}dG_{ij}^{-1})=-dA_i-A_i\wedge A_i
+G_{ij}(dA_j+A_j\wedge A_j)G_{ij}$]. 
In the $U(n)$ case there
was no need to introduce the extra data of the curvature $F$ because at the Lie
algebra level $\Lie(U(n))=Lie(U(n)/U(1))\otimes \Lie(U(1))$,
so that $F$ was canonically constructed from $A$.
\sk

The notion of twisted 1-gerbe (2-gerbe module) can be introduced
performing a similar construction. 
While twisted nonabelian bundles are described by nonabelian
transition functions $\{G_{ij}\}$, twisted nonabelian gerbes are described by 
transition functions $\{f_{ijk}, \varphi_{ij}\}$ that are respectively
valued in $G$ and in  $Aut(G)$, $f_{ijk}: O_{ijk}\to G$, $\varphi_{ij}
:O_{ij}\to Aut(G)$, and where the action of 
$\varphi_{ij}$ on $U(1)$ is trivial: $\varphi_{ij}|_{U(1)}=id$.
The twisted cocycle relations now read 
\eq
\label{twcc0}
\lambda_{ijkl}= 
f_{ikl}^{-1}f_{ijk}^{-1}\varphi_{ij}(f_{jkl})f_{ijl}\,,
\en
\eq
\varphi_{ij}\varphi_{jk}= Ad_{f_{ijk}}\varphi_{ik}\,\label{vphi0}\,,
\en
where $\{\lambda_{ijkl}\}$ is $U(1)$-valued. 
It is not difficult to check that $\{\lambda_{ijkl}\}$ is a 
\v Cech 3-cocycle and thus defines a 2-gerbe (without curvings).
This cocycle may not satisfy the antisymmetry property in its indices,
this however can always be achieved by a gauge transformation with a trivial
cocycle. In the particular case $\lambda_{ijkl}=1$ equations
(\ref{twcc0}), (\ref{vphi0}) define a nonabelian 1-gerbe (without curvings).
\sk

One can also consider twisted gerbes with connections 1-forms:
$(f_{ijk}, \varphi_{ij}, a_{ij}, \mathcal{A}_i)$ where 
$a_{ij} \in \mbox{Lie}(G)\otimes\Omega^1 (O_{ij})$, 
$\mathcal{A}_i \in \mbox{Lie}(Aut(G))\otimes\Omega^1 (O_{i})$,
and twisted gerbes with curvings:
\eq
(f_{ijk}, \varphi_{ij}, a_{ij}, \mathcal{A}_i, B_i, d_{ij}, H_i)
\en
where $ B_i, d_{ij}$ are 2-forms and $H_i$ 3-forms, all of them
$\Lie(G)$-valued;
$B_i \in \mbox{Lie}(G)\otimes\Omega^2(O_{i})$,
$d_{ij}\in\mbox{Lie}(G)\otimes \Omega^2(O_{ij})$,
$H_i \in \mbox{Lie}(G)\otimes\Omega^3(O_{i})$.
Before defining a twisted 1-gerbe we need to introduce some more
notation. Given an element 
$X\in{\mlie}(Aut(G))$, we can construct a
map (a 1-cocycle) $T_X: G\to \mbox{Lie}(G)$ in the following way,
$$
T_X(h) \equiv[ h e^{tX}(h^{-1})]~,
$$
where $[ h e^{tX}(h^{-1})]$ is the tangent vector to the curve 
$h e^{tX}(h^{-1})$ at the point $1_G$; if $X$ is inner, i.e.
$X=ad_Y$ with $Y\in\Lie(G)$,   
then $e^{tX}(h^{-1})=e^{tY}h^{-1}e^{-tY}$ and we simply have 
$T_X(h)=T_{ad_Y}(h)=hYh^{-1}-Y$. Given a ${\mlie}(Aut(G))$-valued form ${\mathcal A}$,
we write ${\mathcal A}={\mathcal A}^\rho X^\rho$ where $\{X^\rho\}$ is a basis of
${\mlie}(Aut(G))$. We then define $T_{{\mathcal A}}$ as
\eq
T_{{\mathcal A}}\equiv {\mathcal A}^\rho T_{X^\rho}~.
\en
We use the same notation $T_{{\mathcal A}}$ for the induced map on ${\mlie}(G)$.
Now we extend this map to allow 
$T_{{\mathcal A}}$ to act on a ${\mlie}(G)$-valued form $\eta=\eta^{\al}Y^\al$,
where $\{Y^\al\}$ is a basis of
${\mlie}(G)$, by $T_{{\mathcal A}}(\eta) = \eta^{\al}\wedge 
T_{{\mathcal A}}(Y^\al)$.
Also we define
\eqa
\mathcal{K}_i &\equiv& d\mathcal{A}_i + \mathcal{A}_i\wedge \mathcal{A}_i\,,\label{K1}\\
k_{ij} &\equiv& da_{ij} + a_{ij}\wedge a_{ij} + T_{\mathcal{A}_i}(a_{ij})\label{k1}\,.
\ena
\sk
%
%
\noi  A twisted 1-gerbe is a set 
$(f_{ijk}, \varphi_{ij}, a_{ij}, \mathcal{A}_i,
 B_i, d_{ij}, H_i)$ such that, $ \varphi_{ij}|_{U(1)}=id,\, 
T_{\!\mathcal{A}_i}|_{U(1)}=0,$
\eq
\varphi_{ij}\varphi_{jk}= Ad_{f_{ijk}}\varphi_{ik}\,,
\label{vphi}
\en
\eq
\mathcal{A}_i + ad_{a_{ij}} = 
\varphi_{ij}\mathcal{A}_{j}\varphi_{ij}^{-1} + \varphi_{ij}d
\varphi_{ij}^{-1}\,,\label{Acc}
\en
\eq
d_{ij} + \varphi_{ij}(d_{jk}) = f_{ijk}d_{ik}f_{ijk}^{-1} +
T_{{\mathcal K}_i+ad_{B_i}}(f_{ijk})\,,\label{deltacc}
\en
\eq
\varphi_{ij}(H_j) = H_i + d\,d_{ij} + [a_{ij},d_{ij}] +
T_{\mathcal{K}_i+ad_{B_i}}(a_{ij}) - T_{\mathcal{A}_i}
(d_{ij}) \,,\label{Hcc}
\en
\sk
\noi and such that 
$
{\mathbf D}_H (f_{ijk}, \varphi_{ij}, a_{ij}, \mathcal{A}_i,
 B_i, d_{ij})\equiv(\la_{ijkl},\al_{ijk},\be_{ij},\ga_i)
$
has  $U(1)$- and $\Lie(U(1))$-valued elements, where
\eqa
\lambda_{ijkl}&\equiv&
f_{ikl}^{-1}f_{ijk}^{-1}\varphi_{ij}(f_{jkl})f_{ijl}\label{twcc}\,, \\
\alpha_{ijk}&\equiv& a_{ij} + \varphi_{ij}(a_{jk}) - f_{ijk}a_{ik}f^{-1}_{ijk} - 
\label{acc}
f_{ijk}df^{-1}_{ijk} - T_{\mathcal{A}_{i}}(f_{ijk})\,,\\
\beta_{ij}&\equiv&  \varphi_{ij}(B_j) - B_i  - d_{ij} + k_{ij}\,,\label{Bcc}\\
\gamma_i &\equiv& H_i -dB_i + T_{\mathcal{A}_i}(B_i)\label{H}\,,
\ena
and we have used the same notation $\varphi_{ij}$
for the induced map $\varphi_{ij} :O_{ij}\to Aut(\mbox{Lie}(G))$.\footnote{In the special case $G=U(1)$ the ${\mathbf D}_H $ operation 
is equivalent to the usual Deligne coboundary operator, provided we 
change $f\rightarrow f^{-1}, B\rightarrow -B$, ($d\rightarrow -d$,
$H\rightarrow -H$) and set $H=0$.}
\sk
If there is zero on the LHS of equations (\ref{acc}), (\ref{Bcc}), (\ref{H})
and $1$ on the LHS of eq. (\ref{twcc}), equations (\ref{vphi})-(\ref{H})
define a nonabelian gerbe\footnote{A general nonabelian 1-gerbe 
is defined by equations (\ref{vphi})-(\ref{H}), where now the group $G$
is an arbitrary group (not necessarily a central
extension).}.
A little algebra, see the appendix, shows that in the less trivial
situation, when we assume that $\lambda_{ijkl}$ is $U(1)$-valued
and $\alpha_{ijk}$, $\beta_{ij}$ and $\gamma_i$ are $\Lie(U(1))$-valued,
the above equations guarantee that
$(\lambda_{ijkl}, \alpha_{ijk}, \beta_{ij}, \gamma_i)$ is a honest
2-gerbe; hence the name twisted 1-gerbe for the set 
$(f_{ijk}, \varphi_{ij}, a_{ij}, \mathcal{A}_i, B_i, d_{ij}, H_i)$.

The 2-gerbe may not satisfy the antisymmetry property in its indices.
This however can always be achieved by a gauge transformation with a trivial
Deligne class.
 
We say that the nonabelian gerbe $(f_{ijk}, \varphi_{ij}, a_{ij}, d_{ij}, A_i,
B_i, H_i)$ is twisted by the 2-gerbe $(\lambda_{ijkl}, \alpha_{ijk},
\beta_{ij},\gamma_i)$. 
We can also say (compare to the one degree
lower situation) that we have a 2-gerbe module,
or that we have a lifting 2-gerbe. The name ``lifting 2-gerbe'' comes from the
following observation:
under the projection $\pi$, that enters the 
group extension
$
U(1)\to G\stackrel{\pi}{\to} G/U(1)\,,
$
the twisting 2-gerbe disappears and we are left with
an ordinary $G/U(1)$-nonabelian gerbe (for example the map $\varphi_{ij}$
is now given by $\pi(\vphi(\hat g))$ and is independent from the
lifting $\hat g$ of the element $g\in G/U(1)$). 
The twisting 2-gerbe is the obstruction
to lift the nonabelian $G/U(1)$-gerbe to a $G$-gerbe.

\sk
\sk
\subsection{Twisted \mbox{\boldmath $\tilde \Omega E_8$} Gerbes}
Consider the exact sequence of groups,
\eq
1\rightarrow \Omega E_8\rightarrow PE_8\,{\stackrel{\pi}{\to}}\, 
E_8\rightarrow 1\label{LEexact}\,,
\en
where  the loop group $\Omega E_8$ is the space of
loops based at the identity $1_{E_8}$, and the based path group $PE_8$ 
is the space of paths starting at the identity
$1_{E_8}$.\footnote{More precisely we should use smooth loops and
paths with sitting instant \cite{Picken}.} Expression
(\ref{LEexact}) states that $\Omega E_8$ is a normal subgroup of $PE_8$,
the quotient being $E_8.$ Consider now the problem of lifting an $E_8$
bundle to a $PE_8$ bundle. Since every path can be homotopically
deformed to the identity path, we have that $PE_8$ is contractible, 
and therefore every $PE_8$ bundle is the trivial bundle. This implies
that only the trivial $E_8$ bundle can be lifted. Any nontrivial $E_8$
bundle cannot be lifted and we thus obtain a nontrivial $\Omega E_8$ 1-gerbe. 
If $\mbox{dim}M \leq 15$ (equivalence classes of) $E_8$ bundles are in 
1-1 correspondence with elements $a\in H^4(M,\BZ)$ and we can say that 
$a$ is the obstruction to lift the $E_8$ bundle, i.e. that $a$
characterizes the gerbe. 
More explicitly the $\Omega E_8$ gerbe has $Aut(\Omega E_8)$ valued maps $\varphi_{ij}$
coming from the conjugation action by some $G_{ij} \in PE_8$ (these
are the transition functions of the twisted $PE_8$ bundle associated
with the $\Omega E_8$ gerbe). Since $\Omega E_8$ is normal in $PE_8$ also the
actions $T_{\mathcal{A}_i}$ of the $\mbox{Lie}(Aut(\Omega E_8))$ valued one 
forms $\mathcal{A}_i$ on $h\in \Omega E_8$ and $X\in \mbox{Lie}(\Omega E_8)$
can be understood as $T_{\mathcal{A}_i}(h) =hA_ih^{-1} - A_i$ and 
$T_{\mathcal{A}_i}(X)= [X,A_i]$ with some 
$\mbox{Lie}(PE_8)$-valued forms $A_i$ that are a lift of the 
connection on the $E_8$ bundle. (See \cite{ACJ} for more details on
gerbes from an exact sequence 
$1\rightarrow H\rightarrow G{\stackrel{\pi}{\to}} G/H\rightarrow 1$).

Finally consider the (universal) central extension of $\Omega E_8$,
\eq
1\rightarrow U(1)\rightarrow \tilde \Omega E_8\,{\stackrel{\pi}{\to}}\, 
\Omega E_8\rightarrow 1\,,
\en
and try to lift the $\Omega E_8$ gerbe to an 
$\tilde \Omega E_8$ gerbe, this is in general not possible and the
obstruction gives
rise to a twisted $\tilde \Omega E_8$ gerbe, the twist being described by a 
2-gerbe. Actually one has {always} an obstruction in
lifting the $\Omega E_8$ gerbe if at least $M\leq 14$, and 
therefore the lifting 2-gerbe thus obtained has characteristic class $a$ 
(characterizing the initial $E_8$ bundle).
The twisted  $\tilde \Omega E_8$ gerbe has $Aut(\tilde \Omega E_8)$ valued maps
$\varphi_{ij}$, obtained extending the previous $\varphi_{ij}$ maps
in such a way that they act trivially on the center $U(1)$ of 
$\tilde \Omega E_8$, also the $T_{\mathcal{A}_i}$ map is 
similarly extended. 

A similar statement holds for $E_8$ replaced by 
$G_2$, $Spin_n$ where $ n\geq 7$, $F_4$, $E_6$, $E_7$, 
when one correspondingly lowers the dimension of $M$.

\sk
\sk
\sk
\section{M5-Brane Anomaly, 
2-Gerbes and Twisted Nonabelian 1-Gerbes}
\label{sec }
In Section 3 cancellation of global anomalies appearing in the open string
worldsheet with strings ending on a stack of D-branes led to condition
(\ref{HW3}) for the D-brane configuration (charges). Here one could in
principle follow a similar approach and study global anomalies of the
path integral of open M2-branes ending on M5
branes.
An alternative 
approach is to study anomalies of 11-dimensional supergravity in
the presence of  M5-branes. The relevant mechanism is the 
cancellation between anomalies of the M5 brane quantum effective action 
and anomaly inflow from the 11-dimensional bulk through a non
invariance of the Chern-Simons plus Green-Schwarz topological term 
$\Phi(C)\sim\int {1\over 6} C\wedge G\wedge G- CI_8(g)$ 
where $C$ is the 3-form potential of 11-dimensional supergravity, $G=dC$ and $I_8\sim (\Tr R^2)^2
-\Tr R^4$, with $R$ being the curvature. We are
interested in the global aspects of this mechanism, where we cannot
assume that $C$ is globally defined and that $G$ is topologically
trivial.
This problem has been studied in \cite{Witten:1996md, Witten:1996hc, 
Witten:1999vg}; and in the more
general case where the 11-dimensional space has boundaries in
\cite{DFM}.
Let $Y$  be the 11-dimensional spacetime: a spin manifold. Let also 
$V$ be the six dimensional M5-brane worldvolume embedded in $Y$
$\iota: V \hookrightarrow Y$, we assume it compact and oriented.
It turns out \cite{Witten:1999vg} that if the field strength $G$
is cohomologically trivial on $V$ and $V$ is the product space 
$V=S\times Q$, with $S$ a circle
with supersymmetric spin structure and $Q$ a five manifold, then the M5-brane
can wrap $V$ iff $Q$ is a Spin$^c$ manifold. If this is not the case
the M5-brane has a global anomaly: one detects it from the vanishing
of the M5-brane partition function. The partition function is zero
every time that there is a torsion element 
$\theta\in H_{tors}^3(Q,\BZ)\subset H^3(Q,\BZ)$ 
different from zero.
More in general, without assuming that $V=S\times Q$, we have a global 
anomaly if there exists an element $\theta\in H_{tors}^4(V,\BZ)$
different from zero.
As suggested in \cite{Witten:1999vg} a way to cancel this anomaly is 
to turn on a background field $G$ such that, essentially,
\eq
\label{anomalyGT}
[G]|_V=\theta
\en
where $[G]|_V$ is the integral class associated with $G$
restricted to $V$.
This condition should be compared to (\ref{HW3}) when the LHS of
(\ref{HW3}) is zero: $[H]|_Q=W_3$. In the case $V=S\times Q$, 
dimensional reduction of the M5-brane on the circle $S$ leads to a
Type IIA D4-brane wrapping $Q$ and satisfying $[H]|_Q=W_3$.

In \cite{DFM} condition (\ref{anomalyGT}) is sharpened.
First a mathematically precise definition of $C$ and of 
$\Phi(C)$ is given, it is in terms of connections on $E_8$ 
bundles. Associated with the field strength $G$ on spacetime
$Y$ with metric $g$, 
there is an integral cohomology class $a\in H^4(Y,\BZ)$. This 
determines an $E_8$ bundle $P(a)\rightarrow Y$ [cf. Section 2.3]. 
The field $C$ can then be described by a couple $(A,c)$ where $A$ is an
$E_8$ connection on $P(a)$ and $c$ is a globally defined
$Lie(U(1))$-valued 3-form on $Y$.  We denote by $\check C=(A,c)$ this $E_8$
description of the $C$-field.  
In particular the holonomy of $\check C$ around a 3-cycle 
$\Sigma$ is given as
$$
\mbox{hol}_\Sigma(\check C)=\mbox{exp}\left[\left(\int_{\Sigma}CS(A) -
\frac{1}{2}CS(\omega) +c\right) \right]\,,
$$
with properly normalized Chern-Simons terms corresponding to the gauge field
$A$ and the spin connection  $\omega$ such that 
$\mbox{exp}[(\int_{\Sigma}CS(A)]$ is well
defined and $\mbox{exp}[{1\over 2}(\int_{\Sigma}CS(\omega)]$ 
has a sign ambiguity. To be more precise these should be the holonomy
of the $E_8$ Chern-Simons 2-gerbe and the proper square root of the
holonomy of the Chern-Simons 2-gerbe associated with the metric.

Subsequently in \cite{DFM}  the electric charge associated with the $C$ field is studied.
From the $C$ field equation of motion that are nonlinear,
$d\star G={1\over 2}G^2-I_8$, we have that the $C$ field and the
background metric induce an electric charge that is given by 
the cohomology class 
\eq
[\frac{1}{2}G^2-I_8]_{DR}\in H^8(Y, \BR)\,. \label{drcharge}
\en
However the electric charge is an integer cohomology class (because
of Dirac quantization, due to the existence of fundamental electric 
M2-branes and magnetic M5-membranes). 
In \cite{DFM} the integral lift of (\ref{drcharge}) 
is studied and denoted $\Theta_Y(\check C)$ (and also $\Theta_Y(a)$).

In order to study the anomaly inflow, we
consider a tubular neighbourhood of $V$ in Y. Since this is diffeomorphic to 
the total space of the normal bundle $N\rightarrow V$, we identify
these two spaces. Let $X=S_r(N)$ be the 10-dimensional sphere bundle of
radius $r$; the fibres of $X {\stackrel{\pi}{\to}}  V$ are then 4-speres. 
An 11-dimensional manifold $Y_r$ with boundary $X$ is
then constructed by removing from $Y$ the disc bundle $D_r(N)$ of radius $r$; 
$Y_r = Y-D_r(N)$ (we can also say that $Y_r$ is the complement of the
tubular neighbourhood $D_r(N)$). We call $Y_r$ the bulk manifold.
Then one has the bulk $C$ field path integral 
$\Psi_{{\!bulk}}(\check C_X)\sim \int
exp[G\wedge\star G] \Phi(\check C_{Y_r})$ where the integral is over all
equivalence classes of $\check C_{Y_r}$ fields that on the boundary assume
the fixed value $\check C_X$. The wavefunction $\Psi_{{\!bulk}}(\check
C_X)$  
is section of a line bundle $\mathcal L$ on the space of $\check C_X$ fields. 
This wavefunction appears together with the M5-brane partition
function $\Psi_{M5}(\check C_V)$ that depends on the $\check C$ field on the
M5-brane, or better, on an infinitesimally small ($r\rightarrow 0$) tubular 
neighbourhood of the M5-brane.
Anomaly cancellation requires $\Psi_{{\!bulk}}\Psi_{M5}$ to be
gauge invariant and therefore $\Psi_{M5}$ has to be a section of the
opposite bundle of $\mathcal L\,$ [$\check C$ fields on $V$ and
$\check C$ fields on
$X$ can be related according to the exact sequence (\ref{Gysin})]. 
Let's study the various cases. 
\sk
I) We can have $\Psi_{{\!bulk}}$ gauge
invariant, and this is shown to imply $\Theta_Y(\check C_X)=0$.  This
last condition is the decoupling condition, indeed if $\Theta_Y(\check
C_X)\not= 0$ then charge conservation requires that M2-branes end on
the M5-brane and the M5-brane is thus not decoupled from the bulk.
If  $\Psi_{{\!bulk}}$ is gauge invariant also  $\Psi_{M5}$ needs to
be, and this holds if $\theta=0$. 
\sk
II) More generally we can have
$\theta\not=0$ but then invariance of $\Psi_{{\!bulk}}\Psi_{M5}$ 
can be shown to imply  
\eq\label{anomcanc}
\pi_*(\Theta_X)=\theta\,,
\en
where $\pi_*$ is integration over the fibre. The map $\pi_*$  
enters the exact sequence
\eq
\label{Gysin}
0\rightarrow H^k(V,\BZ)\stackrel{\pi^*}{\rightarrow}
 H^k(X,\BZ)\stackrel{\pi_*}{\rightarrow} H^{k-4}(V,\BZ)\rightarrow 0\,,
\en
where $\pi^*$ is just pull back associated with the bundle
$X\stackrel{\pi}{\rightarrow}V$. The exactness of this sequence
(obtained from the Gysin sequence) follows from $X$ being oriented 
compact and spin, and $V$ oriented and compact.
Condition (\ref{anomcanc}) is the precise version of condition 
(\ref{anomalyGT}).

We now compare this situation to that in
10 dimensional Type IIA theory, described at the end of Section 3,  
and therefore we are led to consider the following more general case.
\sk
III) Here $\Psi_{{\!bulk}}\Psi_{M5}$ is not gauge invariant
(therefore it is a section of a line bundle) but we can consider 
a new partition function 
$\Psi'_{M5}$ that is obtained from a ``stack'' of M5-branes instead of just
a single brane. This stack gives rise to a twisted gerbe 
$(f_{ijk}, \varphi_{ij}, a_{ij}, \mathcal{A}_i,
 B_i, d_{ij}, H_i)$
on $V$ so that in particular $\Psi'_{M5}$ depends also from the nonabelian 
gauge fields $B_i$ and $H_i$\footnote{Of course we have the special
  case when a stack of M5-branes gives a nonabelian gerbe. Then $H_i$
  is the curvature of $B_i$. These two fields should not be confused, and
  have nothing to do with the NS $B$ field and its curvature $H$.}. 
In order for
$\Psi_{{\!bulk}}\Psi_{M5}'$ to be well defined, 
the twisted gerbe has to satisfy [cf. (\ref{abcde})],
\eq\label{gentwist}
[CS({\pi_*(\Theta_X)})]-[\vartheta_{ijkl},0,0,0]=[
{\mathbf D}_H (f_{ijk}, \varphi_{ij}, a_{ij}, \mathcal{A}_i,
 B_i, d_{ij})]+[1,0,0,C_V]\,,
\en
where, as constructed in Subsection 2.3,  
$CS({\pi_*(\Theta_X)})$ is the Chern-Simons 2-gerbe associated
with $\pi_*(\Theta_X)$ and a choice 
of connection on the $E_8$ bundle with first Pontryagin class 
${\pi_*(\Theta_X)}$ (all other 2-gerbes 
differ by a global 3-form, see (\ref{sequence1})),
while $[\vartheta_{ijkl},0,0,0]$ is the 2-gerbe class
associated with the torsion class $\theta$
[i.e. $\beta(\vartheta)=\theta$, cf (\ref{24})], and $[1,0,0,C_V]$
is the trivial Deligne class associated with the global 3-form $C_V$.

In particular (\ref{gentwist}) implies
\eq
\pi_*(\Theta_X)-\theta=
\xi_{{\mathbf D}_H (G_{ijk}, \varphi_{ij}, a_{ij}, \mathcal{A}_i,
 B_i, d_{ij})}\,,
\en
where on the RHS we have the characteristic class of the lifting 2-gerbe.
\sk
The correspondence of this construction with that described in Section
3, is strengthened by slightly generalizing the results of Section 3.
In fact there we always considered $[H]|_Q-W_3$ 
trivial in De Rham cohomology. This implied that the 
torsion class $[H]|_Q-W_3$ was interpreted as the characteristic class
of a gerbe associated with a twisted $U(n)$ bundle for some $n\in
\BZ$.
However (at least mathematically) one can consider the more general 
case where $[H]|_Q-W_3\not=0$ also in De Rham cohomology. Here too
we have a twisted bundle, but with structure group $U({\HH})$, the
group of unitary operators on the complex, separable and infinite
dimensional Hilbert space $\HH$. This case corresponds to an infinite
number of D-branes wrapping the cycle $Q$, and the relevant central
extension is $U(1)\rightarrow  U(\HH)\rightarrow PU(\HH)$.
When $\mbox{dim}Q \leq 13$ (which is always the case in
superstring theory), we can replace, for homotopy purposes,  
$PU(\HH)$ with $\Omega E_8$ and $U(\HH)$ with $\tilde \Omega E_8$, 
so that the group extension
$U(1)\rightarrow  U(\HH)\rightarrow PU(\HH)$ is replaced with
$U(1)\rightarrow  \tilde\Omega E_8\rightarrow \Omega E_8$.
Now consider a stack of M5-branes wrapping a cycle $V=S\times Q$ 
and dimensionally reduce M-theory to Type IIA
along the circle $S$. Then the M5-branes become
D4-branes and the twisted $\Omega E_8$ 
1-gerbe becomes a twisted $\Omega E_8$ bundle.
\sk\sk
\sk

\begin{center}
\textsc{Acknowledgements}
\end{center}
We would like to thank Peter Bouwknegt, Dale Husemoller, Jouko Mickelsson
and Stefan Theisen for useful discussions. P.~A. would like to thank
Masud Chaichian for the hospitality at University of 
Helsinki during completion of this work, and the financial 
support of the Academy of Finland under Projects 
No.~54023 and 104368.

\sk

\sk
\sk
\appendix


\section{Proof that a twisted 1-gerbe defines a 2-gerbe}

The cocycle condition for $\la_{ijkl}$ is straightforward.
In order to show that $\alpha_{ijk}$ as defined in (\ref{acc}) 
satisfies the 2-gerbe condition
$$
\al_{ijk} + \alpha_{ikl} - \alpha_{ijl} - \alpha_{jkl} =
\lambda_{ijkl}d\lambda_{ijkl}^{-1}\,,
$$
we rewrite the LHS as
$
\alpha_{ijk} + Ad_{f_{ijk}}\alpha_{ikl} - 
Ad_{\varphi_{ij}(f_{jkl})}\alpha_{ijl} - \alpha_{jkl}\,,
$
we then use the definition of $\alpha_{ijk}$ and the following properties
of the map  $T_\AAA$,
\eqa
T_\AAA(hk)&=&T_\AAA(h)+kT_\AAA(h)k^{-1}\,, ~~~~~~~~~~~{\mbox{ (cocycle property)}}\\
\vphi_{ij}(T_\AAA(h))&=&T_{\vphi_{ij}\AAA\vphi_{ij}^{-1}}(\vphi_{ij}(h))\,,\\
T_{-\vphi_{ij}d\vphi_{ij}^{-1}}(\vphi_{ij}(h))&=&
\vphi_{ij}(h)d\vphi_{ij}(h^{-1})-
\vphi_{ij}(hdh^{-1})\,,
\ena
where $h,k$ are elements of $G$, and more in general functions 
from some open neighbourhood of $M$ into $G$. 
Finally  $T_{\AAA_i}(\varphi_{ij}(f_{jkl})f_{ijl})=
T_{\AAA_i}(f_{ijk}f_{ikl})$ since $T_{\AAA_i}(\la_{ijkl})=0$.

Similarly in order to show that 
$$
\beta_{ij}+ \beta_{jk} + \beta_{ki} = d \alpha_{ijk} 
$$
we rewrite the LHS as $\beta_{ij}+ \vphi_{ij}(\beta_{jk}) +
f_{ijk}\beta_{ki}f_{ijk}^{-1}$ and then use 
the following equality
\eq
k_{ij} + \varphi_{ij}(k_{jk}) = f_{ijk}k_{ik}f_{ijk}^{-1} + T_{\mathcal{K}_i}(f_{ijk}) + d\alpha_{ijk}\,,
\en
that follows from (\ref{K1})-(\ref{Acc}), the algebra here is the same
as for usual gerbes. We also have 
$
\mathcal{K}_i + ad_{k_{ij}} = \varphi_{ij}\mathcal{K}_j\varphi_{ij}^{-1}\,,
$
and the Bianchi identity 
\eq
dk_{ij}+[a_{ij},k_{ij}]+T_{\mathcal{K}_i}(a_{ij})-
T_{\mathcal{A}_i}(k_{ij})\,.\label{Binent}
\en
Relation 
$$
\gamma_i - \gamma_j = d \beta_{ij}
$$
that we rewrite as $\gamma_i - \vphi_{ij}(\gamma_j) = d \beta_{ij}$
follows from (\ref{Binent}), (\ref{Hcc}) and 
$
T_{-\vphi_{ij}d\vphi_{ij}^{-1}}(\vphi_{ij}(B_j))=
-d(\vphi_{ij}(B_j))+
\vphi_{ij}(dB_j)\,.
$
\sk\sk

\sk\sk


\begin{thebibliography}{33}
\raggedright


\bibitem{Witten:1998cd}
E.~Witten,
``D-branes and K-theory,''
JHEP {\bf 9812} (1998) 019
[arXiv:hep-th/9810188]


\bibitem{Kapustin:1999di}
A.~Kapustin,
``D-branes in a topologically nontrivial B-field,''
Adv.\ Theor.\ Math.\ Phys.\  {\bf 4}, 127 (2000)
[arXiv:hep-th/9909089]

\bibitem{Bouwknegt:2000qt}
P.~Bouwknegt and V.~Mathai,
``D-branes, B-fields and twisted K-theory,''
JHEP {\bf 0003} (2000) 007
[arXiv:hep-th/0002023]


\bibitem{Bouwknegt:2001vu}
P.~Bouwknegt, A.~L.~Carey, V.~Mathai, M.~K.~Murray and D.~Stevenson,
``Twisted K-theory and K-theory of bundle gerbes,''
Commun.\ Math.\ Phys.\  {\bf 228}, 17 (2002)
[arXiv:hep-th/0106194]


\bibitem{Carey:2002xp}
A.~L.~Carey, S.~Johnson and M.~K.~Murray,
``Holonomy on D-branes,''
arXiv:hep-th/0204199


\bibitem{Witten:1996hc}
E.~Witten,
``Five-brane effective action in M-theory,''
J.\ Geom.\ Phys.\  {\bf 22} (1997) 103
[arXiv:hep-th/9610234]


\bibitem{Witten:1999vg}
E.~Witten,
``Duality relations among topological effects in string theory,''
JHEP {\bf 0005} (2000) 031
[arXiv:hep-th/9912086]



\bibitem{DFM}
E.~Diaconescu, G.~W.~Moore and D.~S.~Freed,
``The M-theory 3-form and E(8) gauge theory,''
arXiv:hep-th/0312069


\bibitem{Moore:2004jv}
G.~W.~Moore,
``Anomalies, Gauss laws, and page charges in M-theory,''
arXiv:hep-th/0409158

\bibitem{Hopkins:2002rd}
M.~J.~Hopkins and I.~M.~Singer,
``Quadratic functions in geometry, topology, and M-theory,''
arXiv:math.at/0211216


\bibitem{Breen}
L.~Breen, W.~Messing
`` Differential Geometry of Gerbes,''
arXiv:math.AG/0106083


\bibitem{ACJ}
P.~Aschieri, L.~Cantini and B.~Jurco,
``Nonabelian bundle gerbes, their differential geometry and gauge theory,''
To appear in Comm. Math. Phys.,
arXiv:hep-th/0312154


\bibitem{Witten:1996md}
E.~Witten,
``On flux quantization in M-theory and the effective action,''
J.\ Geom.\ Phys.\  {\bf 22}, 1 (1997)
[arXiv:hep-th/9609122]



\bibitem{DMW}
D.~E.~Diaconescu, G.~W.~Moore and E.~Witten,
``A derivation of K-theory from M-theory,''
arXiv:hep-th/0005091


\bibitem{Evslin}
A.~Adams and J.~Evslin,
``The loop group of E(8) and K-theory from 11d,''
JHEP {\bf 0302} (2003) 029
[arXiv:hep-th/0203218]

\bibitem{Sati}
V.~Mathai and H.~Sati,
``Some relations between twisted K-theory and E(8) gauge theory,''
JHEP {\bf 0403}, 016 (2004)
[arXiv:hep-th/0312033]

\bibitem{HW}
P.~Ho\v rava and E.~Witten,
``Eleven-Dimensional Supergravity on a Manifold with Boundary,''
Nucl.\ Phys.\ B {\bf 475}, 94 (1996)
[arXiv:hep-th/9603142]


\bibitem{West}
P.~C.~West,
``E(11) and M theory,''
Class.\ Quant.\ Grav.\  {\bf 18} (2001) 4443
[arXiv:hep-th/0104081]

\bibitem{D'Auria:1982nx}
R.~D'Auria and P.~Fr\'e,
``Geometric Supergravity In D = 11 And Its Hidden Supergroup,''
Nucl.\ Phys.\ B {\bf 201} (1982) 101
[Erratum-ibid.\ B {\bf 206} (1982) 496]


\bibitem{Bandos}
I.~A.~Bandos, J.~A.~de Azcarraga, M.~Picon and O.~Varela,
``On the formulation of D = 11 supergravity and the composite nature
of its three-form field,''
arXiv:hep-th/0409100


\bibitem{Hitchin}
N.~Hitchin,
``Lectures on special Lagrangian submanifolds,''
arXiv:math.dg/9907034


\bibitem{Brylinski}
J.~L.~Brylinski,
``Loop Spaces, Characteristic Classes And Geometric Quantization,''
Progress in mathematics  {\bf{107}},
Birkh\"auser, Boston (1993) 

\bibitem{Cheeger}
J. Cheeger and J. Simons, "Differential characters and geometric invariants", Stony Brook Preprint 1973; reprinted in Lecture Notes in Mathematics 1167, Geometry and Topology Proc. 1983-84, Eds J. Alexander and J. Harer, Springer 1985


\bibitem{Freed:1999vc}
D.~S.~Freed and E.~Witten,
``Anomalies in string theory with D-branes,''
arXiv:hep-th/9907189


\bibitem{Stora}
M.~Bauer, G.~Girardi, R.~Stora and F.~Thuillier,
``A class of topological actions,''
arXiv:hep-th/0406221

\bibitem{Gajer}
P. Gajer, ``Geometry of Deligne Cohomology,''
Invent. Math. {\bf 127}, no. 1, 155-207, 1997, [arXiv:alg-geom/9601025]


\bibitem{Jouko}
J. Mickelsson "Current Algebras and Groups,'' Plenum Press, New York, 1989

\bibitem{Johnson}
S. Johnson, `` Constructions with bundle gerbes,''
Ph.D. thesis,  math.DG/0312175
   
     
\bibitem{Witten:1985}
E.~Witten,
``Topological Tools In Ten-Dimensional Physics,''
Int.\ J.\ Mod.\ Phys.\ A {\bf 1} (1986) 39.


\bibitem{Gawedzki}
K.~Gawedzki, ``Topological actions in two-dimensional quantum field
theories'', in nonperturbative quantum field theory, G. 't Hooft et
al. ed. (Cargese 1987) 101-141, NATO Adv. Sci. Inst. Ser. B:
Phys., 185,  Plenum Press, New York, 1988

\noi K.~Gawedzki and N.~Reis,
``WZW branes and gerbes,'' Rev.Math.Phys. {\bf{14}}  \rm1281-1334 (2002)
[arXiv:hep-th/0205233]


\bibitem{G}
Grothendieck, A, ``Le groupe de Brauer'', S\'eminaire Bourbaki,
Vol. 9, exp 290, Soc. Math. France 1995

\bibitem{Picken}
A.~Caetano and R.~F.~Picken,
``An Axiomatic definition of holonomy,''
Int. J. Math., 5(6):835-848, 1994
[IFM-14-93, see also KEK Library]


\bibitem{Murray}
M.~ Murray, ``Bundle Gerbes,''  J. London Math. Soc.,54:403-416, 1996, 
[arXiv:dg-ga/9407015]


\end{thebibliography}
\end{document}